\documentclass[journal=jacsat,manuscript=article]{achemso}
\usepackage[version=4]{mhchem} 
\usepackage[T1]{fontenc} 
\usepackage[separate-uncertainty=true]{siunitx}

\author{Maxence Menétrey}
\email{maxence.menetrey@mat.ethz.ch}
\author{Lukáš Zezulka}
\affiliation[ETH Zurich]{Laboratory for Nanometallurgy, Department of Materials, ETH Zurich, Vladimir-Prelog-Weg 1-5/10, Zürich, 8093, Switzerland}
 \alsoaffiliation[Institute of Physical Engineering]{Institute of Physical Engineering, Faculty of Mechanical Engineering, Brno University of Technology, Brno 616 69, Czech Republic}
\author{Pascal Fandré}
\author{Fabian Schmid}
\author{Ralph Spolenak}
\email{ralph.spolenak@mat.ethz.ch}
\affiliation[ETH Zurich]{Laboratory for Nanometallurgy, Department of Materials, ETH Zurich, Vladimir-Prelog-Weg 1-5/10, Zürich, 8093, Switzerland}
\affiliation[Institute of Physical Engineering]{Institute of Physical Engineering, Faculty of Mechanical Engineering, Brno University of Technology, Brno 616 69, Czech Republic}

\title[Nanodroplet Flight Control in Electrohydrodynamic Redox 3D Printing]
  {Nanodroplet Flight Control in Electrohydrodynamic Redox 3D Printing}

\abbreviations{}

\begin{document}

\begin{abstract}

Electrohydrodynamic 3D printing is an additive manufacturing technique with enormous potential in plasmonics, microelectronics, and sensing applications, thanks to its broad materials palette, high voxel deposition rate, and compatibility with various substrates.
However, the electric field used to deposit material is concentrated at the depositing structure resulting in the focusing of the charged droplets and geometry-dependent landing positions, which complicates the fabrication of complex 3D shapes.
The low level of concordance between design and printout seriously impedes the development of electrohydrodynamic 3D printing and rationalizes the simplicity of the designs reported so far.
In this work, we break the electric field centrosymmetry to study the resulting deviation in the flight trajectory of the droplets.
Comparison of experimental outcomes with predictions of an FEM model provides new insights into the droplet characteristics and unveils how the product of droplet size and charge uniquely governs its kinematics.
From these insights, we develop reliable predictions of the jet trajectory and allow the computation of optimized printing paths counterbalancing the electric field distortion, thereby enabling the fabrication of geometries with unprecedented complexity.

\end{abstract}

\newpage
\section{Introduction}

The autofocusing effect is a double-edged sword for electrohydrodynamic (EHD) 3D printing.
On the one hand, this phenomenon—describing the attraction of charged droplets towards a printed object in response to the electric field concentration it creates—helps the fabrication of high-aspect-ratio structures and ensures sub-\SI{100}{\nano\meter} resolution \cite{Galliker2012}.
It also allows for the precise positioning of quantum dots on plasmonically active structures for photonics applications \cite{Kress2014} and improves stitching reliability in multi-material printing using both, a multi-nozzle \cite{An2015} and a single nozzle with double-channel \cite{Reiser2019} approach.
On the other hand, the non-trivial droplet deflection involved detrimentally complicates the fabrication of 3D architectures.
Compared to localized electrodeposition \cite{ Ercolano2020, Lin2019} or focused electron/ion beam-induced deposition (FEBID/FIBID) \cite{ Winkler2021, Skoric2020}, EHD printing is relatively limited in its geometric freedom due to this autofocusing effect.
Despite the technique's unique and promising advantages—compatibility with various types of substrates, large materials palette, and high deposition rate\cite{Hirt2017}—even fabrication of seemingly simple geometries is impacted.
For instance, printing a wall that intersects a previously printed one leads to undesirable discontinuities in the structure \cite{Schirmer2011}, and this is only partially improved by keeping a low aspect ratio (wall height over thickness) in combination with concomitant printing of the two walls \cite{Schneider2016}.
As further breakthroughs in EHD 3D printing will require an approach for the fabrication of arbitrary shapes, overcoming the challenges of autofocusing is critical to pushing beyond pillars, spirals \cite{An2015}, bridges \cite{An2015}, and walls \cite{Rohner2020} and unleashing the potential of the technique.
Reaching this crucial milestone necessitates, firstly, a better understanding of the droplet dynamics in the EHD process and, secondly, a versatile strategy for creating printing paths that compensate for the autofocusing effect.

To that end, efforts were made to understand how perturbations in the electric field by the deposited voxels (the 3D equivalent of a pixel) influence the droplet dynamics and thus the landing position of the subsequent voxels.
Yudistira \textit{et al.} systematically studied the deflection and the breakup of ejected droplets in response to the accumulation of charged solvent on a dielectric substrate \cite{Yudistira2010}.
In the case of electrospun fibers—another application of EHD printing—the charge carried by the collected fibers strongly influences the deposition of the following ones\cite{Ding2019}.
Through conscientious design, these electrostatic interactions can be harnessed to allow for precise positioning of EHD fibers and droplets.
By introducing external deflection electrodes, charged droplets can be deflected in a controlled way to create macroscale 2D patterns \cite{Plog2020}.
Extending this approach, Liashenko \textit{et al.} demonstrated the electrostatic steering of electrospun fibers to create well-defined 2D patterns and out-of-plane stacks \cite{Liashenko2020}.
The deflection speed enabled by the external electrodes outperforms the capabilities of mechanical stages, and thus allows for fibers alignment without buckling.

In parallel, efforts were made to alleviate the negative impact of the electrostatic deflection.
O'Connell \textit{et al.} derived an empirical model to compute printing paths counteracting the effect of electric field distortion around hemispheres of different sizes \cite{OConnell2021}.
Though successful for this particular geometry, the printing path correction for an arbitrary shape—which continuously evolves during printing—requires a more agile and comprehensive approach, such as numerical simulation by finite element method (FEM).
Other nanoscale printing techniques, such as aerosol jets printing \cite{Jung2021} or FEBID/FIBID \cite{Weitzer2022, Fowlkes2016}, readily benefited from FEM modeling for geometrical optimization, but work in EHD is still emerging.
The modeling of EHD jets assisted in the design of a multi-nozzle array print head that addresses the crosstalk issues between individual nozzles \cite{Duan2022}.
Richner \textit{et al.} provided a more fundamental study by observing the deflection of EHD droplets from a vertical path in response to electric field modification by thin-film electrodes \cite{Richner2016d}.
Their FEM model determined the droplet trajectory and provided—by comparison with experimental results—an estimation for the charge and the size of the ejected droplets.
Despite the increasing interest in numerical methods for understanding the EHD process and improving jet straightness from a multi-nozzle array, it has not yet been employed to optimize geometrical conformity between design and print outcome.

To tackle this challenge of high geometric fidelity in EHD printing, we here develop a combined experimental and modeling approach to investigate the flight dynamics of charged droplets using electrohydrodynamic redox 3D printing (EHDRP) process as a model system.
EHDRP has recently been introduced as an electrochemical alternative to EHD \cite{Reiser2019, Nydegger2022}, in which the nanoparticles suspension is replaced with metal ions in solution. 
This technique advantageously combines both the high deposition rate of EHD and the various benefits of EC-based nanoscale 3D printing \cite{Hengsteler2022}, including superior as-deposited materials quality \cite{Reiser2020}, tunable microstructures \cite{Menetrey2022, Menetrey2023, Behroozfar2018}, and no requirement for post-printing annealing steps.
Notably, EHDRP distinguishes itself from other nanoscale additive techniques by its potential for multi-materials printing \cite{Reiser2019} and its unique ability to synthesize multi-component alloys with precise compositions \cite{Porenta2023}.

Here, the electrostatic deflection of charged droplets using thin-film electrodes allows for their precise positioning to create various geometries.
We also examine the flight path alteration resulting from the presence of a nanopillar printed on the substrate to provide physical basis for the autofocusing effect in EHDRP.
In comparing these experimental observations with predictions of the developed FEM simulations, this work provides new insights into the droplet characteristics and their role in the deflection of the droplet's path.
Finally, our development of an FEM-based optimization tool highlights key insights into the autofocusing phenomena in EHDRP and enables printing paths correction for the electrostatic autofocusing. 
Through these tools, EHDRP and other EHD techniques can reach greater geometric complexity and open the promise for fast, tunabel microstructural design at the nanoscale.

\newpage
\section{Results and discussion}

\begin{figure}[htb]
    \includegraphics[width=16.29cm]{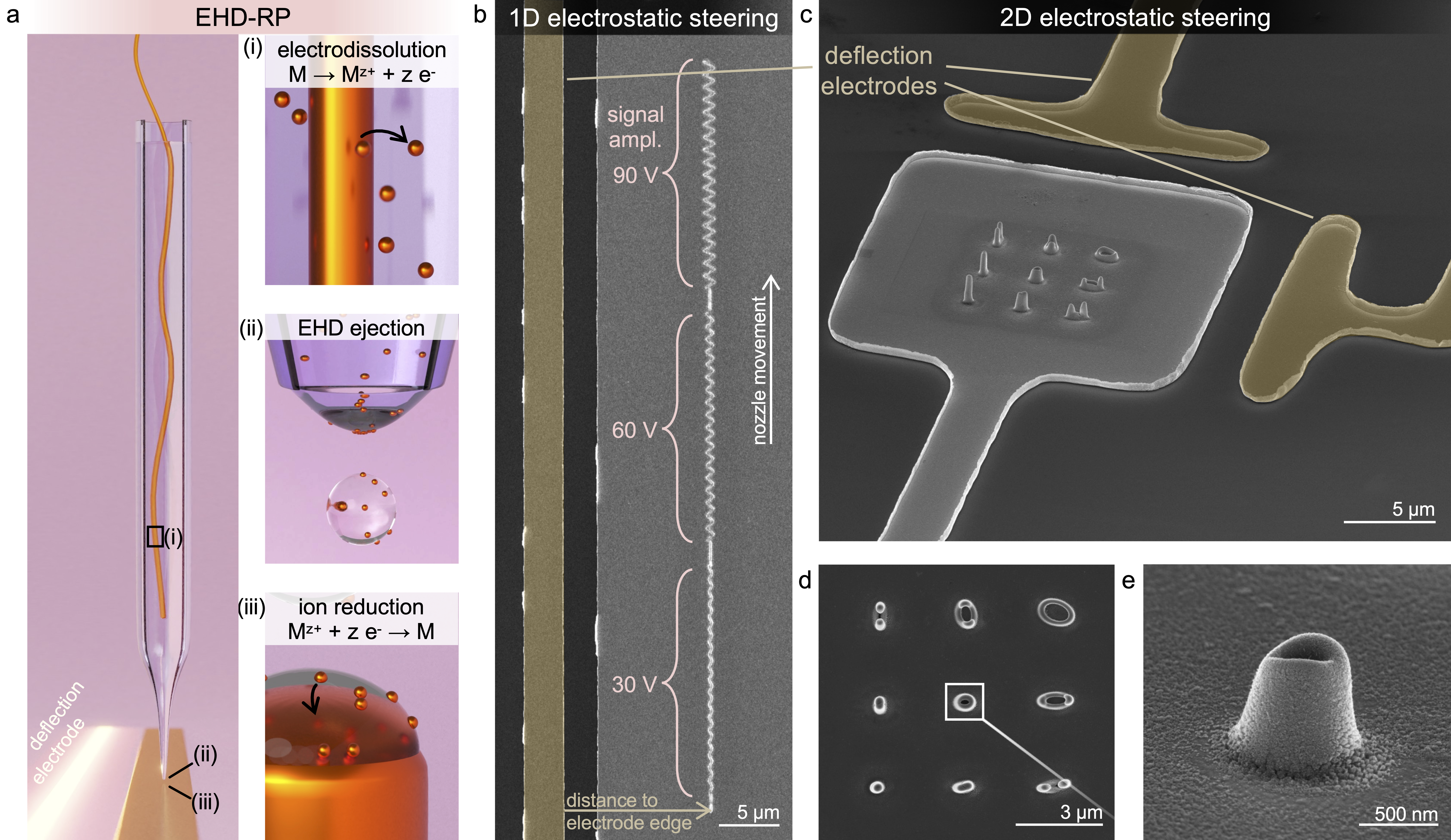}
	\caption{\textbf{EHDRP with electrostatic jet deflection.} (a) Schematic of the EHD process. A sacrificial copper anode inserted in an acetonitrile-filled quartz capillary is polarized with a $\approx\SI{100}{\volt}$ potential. The strongly positive voltage forces (i) the generation of solvated metal ions, while the electric field triggers (ii) the electrohydrodynamic ejection of ion-loaded droplets. Upon landing (iii), the ions are reduced and the solvent evaporates, leading to the creation of a metallic voxel. (b) SE micrographs showing 1D deflection of the EHD jet by an electrode parallel to the nozzle movement and polarized with a \SI{30}{}, \SI{60}{}, and \SI{90}{\volt} voltage amplitude. (c) SE micrographs showing electrostatic deflection of the EHD jet in all in-plane directions using two perpendicular and independently polarized electrodes. (d–e) Out-of-plane cylinders deposited with a constant nozzle position and a sinusoidal signal applied to both deflection electrodes. Imaging tilt: (b,d) 0°; (c,e) 55°.}
	\label{fig:Figure_1}
\end{figure}

\subsection{Electrostatic deflection by external electrodes}

Electrohydrodynamic redox 3D printing (Figure~\ref{fig:Figure_1}(a)) relies on the polarization of an immersed sacrificial anode with a $\approx\SI{100}{\volt}$ electric potential, leading to (i) the electrodissolution of the anode—thereby generating solvated metal ions—and (ii) the electrohydrodynamic ejection of ions-loaded solvent droplets from a quartz capillary nozzle.
Upon landing on a grounded substrate, the solvent evaporates and (iii) the ions are reduced, giving rise to a metallic voxel.
The droplet formed in this process carries a charge; thus, its flight is impacted by the orientation of the electric field.
In a standard EHD configuration where the autofocusing effect is not considered, the electric field arises solely from the potential difference between the nozzle and the counter electrode—e.g., a grounded flat substrate—leading to a spatial configuration with circular symmetry and thus a straight droplet flight path.
In order to study the droplet deflection, we break the circular symmetry of the electric field by introducing polarizable thin-film electrodes in the vicinity of the printing nozzle.
Using only one electrode (1D) with nozzle translation parallel to the electrode (Figure~\ref{fig:Figure_1}(b)), the shape of the deposited copper lines strictly follows the triangular waveform of the electrical signal.
Also, the magnitude of the deflection appears to increase linearly with the signal amplitude (Figure~S1(a)).

Electrostatic deflection offers a unique approach for the precise positioning of voxels with lateral jet speeds and accelerations unattainable by nano-positioning systems.
The introduction of a second electrode perpendicular to the first unlocks complete in-plane positioning of the droplet and enables the deposition of 2.5D structures (Figure~\ref{fig:Figure_1}(c)).
In this approach, a sinusoidal signal applied to both electrodes with a $\pi$ phase shift between the two leads to the deposition of a cylinder (Figure~\ref{fig:Figure_1}(e)).
Interestingly, the 9 cylinders visible here stretch and/or compress along the two principal axes (Figure~\ref{fig:Figure_1}(d)).
Given that each cylinder was fabricated with the same signal amplitude but with 9 different nozzle positions, the strength of the electric field produced by the deflection electrode rapidly decays with the distance.

\begin{figure}[htbp]
    \centering
    \includegraphics[width=17.78cm]{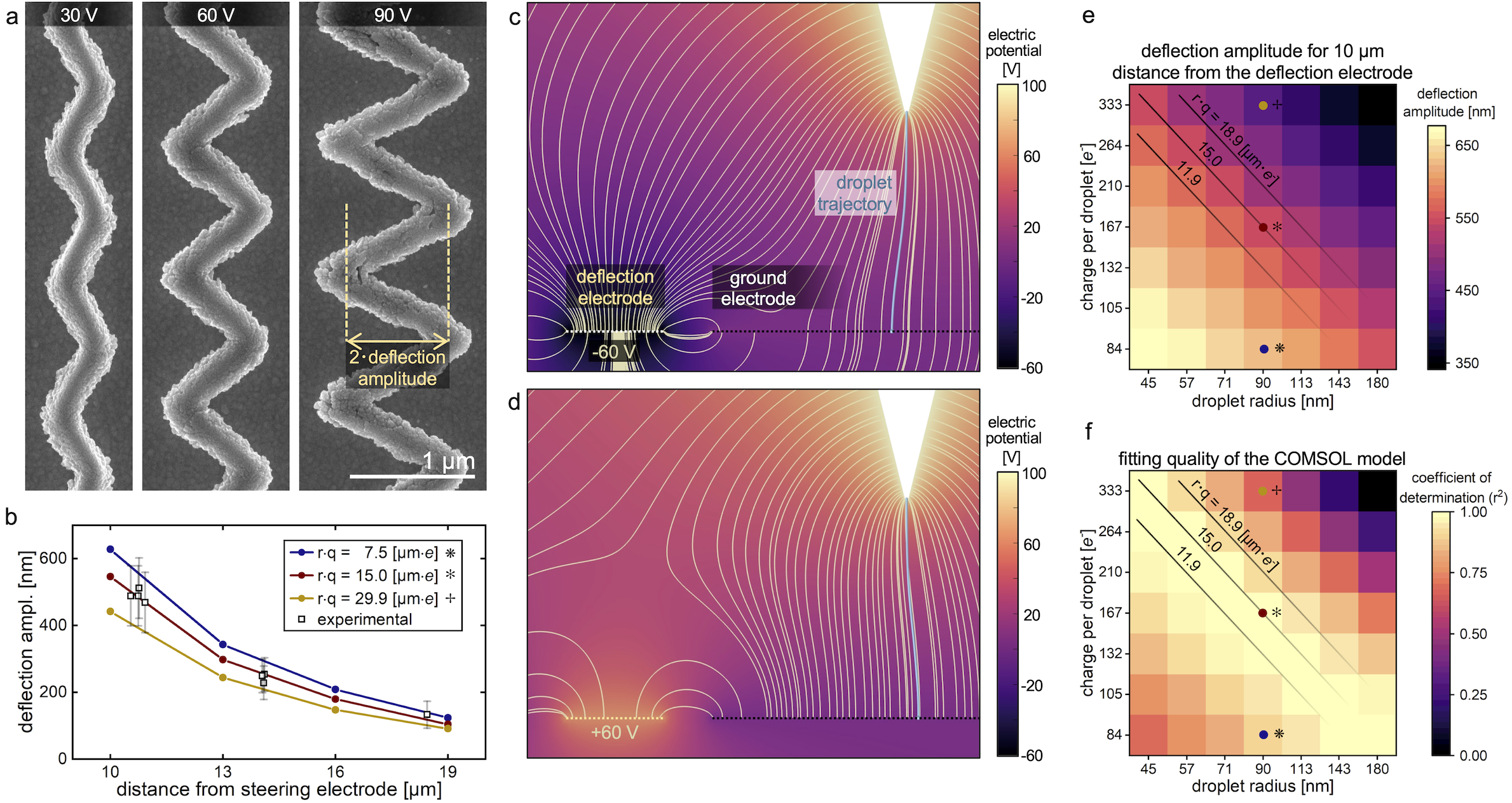}
	\caption{\textbf{Electrostatic deflection and comparison to FEM modeling.} (a) Copper lines showing waviness as a result of electrostatic deflection with a triangular waveform. The amplitude of the deflection directly follows the signal amplitude. (b) Deflection amplitude as a function of the distance from the steering electrode for a \SI{60}{\volt} signal amplitude as obtained experimentally and derived from FEM modeling for droplets with various $radius \cdot charge$ ($r \cdot q$) products. The error bars are an estimation of the confidence interval for the deflection measurement from SE micrographs. (c–d) 2D maps of the electric potential distribution predicted by the 3D FEM COMSOL Multiphysics\textsuperscript{\textregistered} model showing the electric field lines and the simulated droplet path for a deflection voltage of (c) \SI{-60}{} and (d) $+$\SI{60}{\volt}. The droplet in these maps has a $\SI{90}{\nano\meter}$ radius and carries a charge of $\SI{167}{\elementarycharge}$. (e) Model predictions for the deflection amplitude, taken as half the peak-to-peak value between \SI{-60}{} and $+$\SI{60}{\volt} deflection voltage, and (e) coefficient of determination (r\textsuperscript{2}) showing the degree of agreement between the model predictions and experimental data for different droplet radii and number of elementary charges. The results unveil the relationship that links the droplet deflection to its $r \cdot q$ product.}
	\label{fig:Figure_2}
\end{figure}

The post-printing deflection amplitude is derived from SE micrographs in the 1D configuration (Figure~\ref{fig:Figure_2}(a)) as a function of the distance from the steering electrode edge and compared to the deflection amplitude predicted by a 3D FEM model (Figure~\ref{fig:Figure_2}(b)).
The model—built in the COMSOL Multiphysics\textsuperscript{\textregistered} environment—uses the Electrostatics module to predict the electric field distribution and, thereafter, computes the trajectory of a spherical droplet within this field using the Charged Particle Tracing module.
The assumption of droplet sphericity is discussed in Section~S1 (Supporting Information).
In accordance with experimental measurement, the amplitude predicted by the model decreases with the distance from the steering electrode (Figure~\ref{fig:Figure_2}(b)).
A cross-sectional map of the simulated electric potential distribution is displayed in Figure~\ref{fig:Figure_2}(c–d) for a \SI{-60}{} and $+$\SI{60}{\volt} deflection potential. 
In these two cases, the influence of the electrode voltage on both the electric field lines between the nozzle and substrate (light yellow) and the droplet path (light blue) is visible.
The model also shows that electrode potential has little to no impact on the electric field lines in the vicinity of the nozzle opening.
Thus the droplet trajectory does not deviate from a vertical path right after ejection from the nozzle but rather at a later stage.

The amplitude of the deflection from the vertical path appears to be closely linked to the characteristics of the droplet, for instance, its radius and the number of elementary charges it carries (Figure~\ref{fig:Figure_2}(e)).
The model indicates a greater degree of deflection with applied lateral electric field in lighter or more weakly charged droplets.
More specifically, all droplets with matching $radius \cdot charge$ ($r \cdot q$) products undergo a similar degree of deflection.
Accordingly, the best agreement between the FEM model and the experiments (Figure~\ref{fig:Figure_2}(f)) is found for all couples of $r$ and $q$ whose product lies around \SI{12}{}–\SI{15}{\micro \meter \elementarycharge} (or between \SI{7}{}–\SI{30}{\micro \meter \elementarycharge} to stay conservative with respect to the uncertainty in the experimental data).
We further note that the $r \cdot q$ product of the droplet is not significantly altered by a change in the printing voltage from \SI{100}{} to \SI{130}{\volt} (see Figure~S1(b)).
Thus any variation in droplet size with printing voltage would be, in first approximation, compensated by a variation in charge that keeps a constant $r \cdot q$ product.

\subsection{Droplet deflection in response to the autofocusing effect}

\begin{figure}[htbp]
    \centering
    \includegraphics[width=17.78cm]{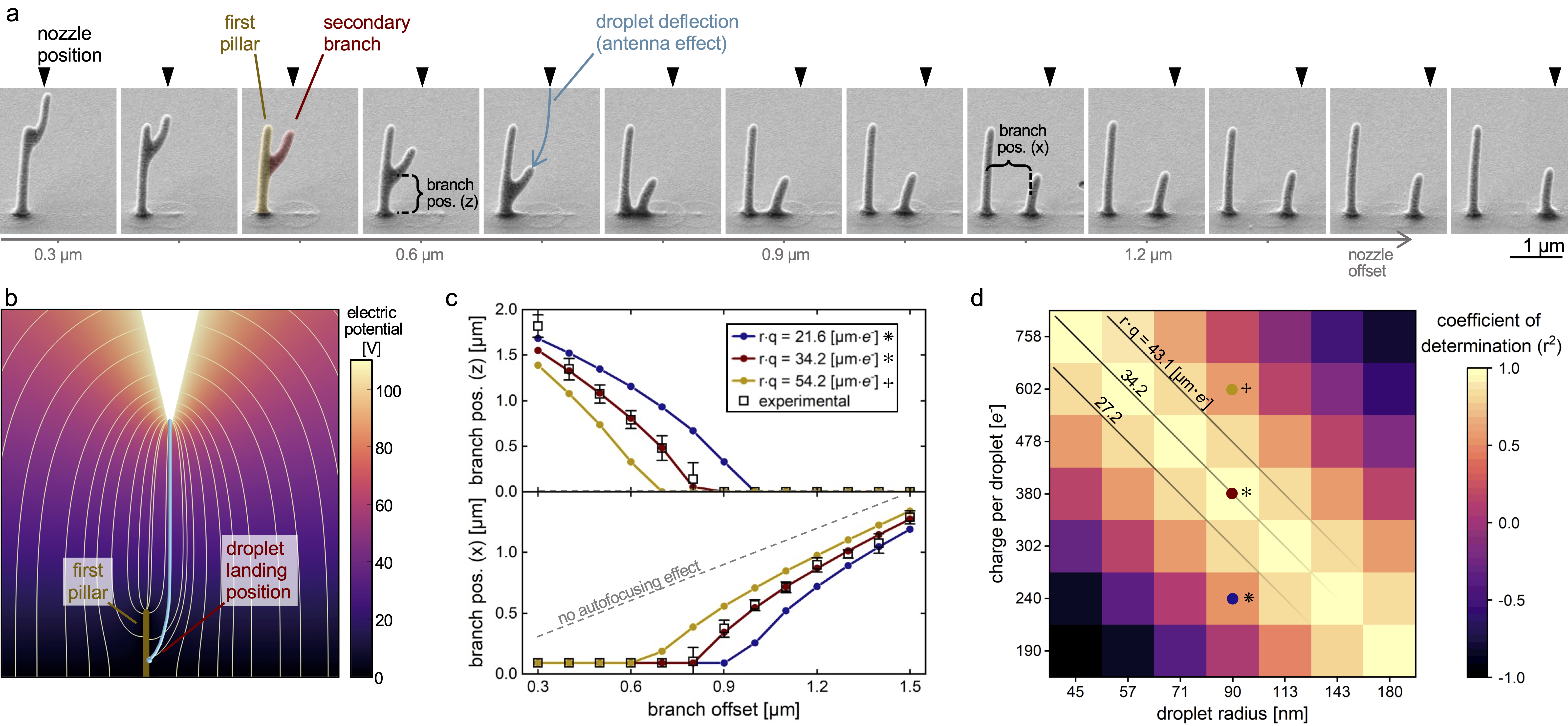}
	\caption{\textbf{Autofocusing effect and predictions of the FEM model.} (a) SE micrographs revealing how the presence of a nanopillar influences the deposition of a subsequent branch. Imaging tilt: 55°. (b) 2D cross-section of the electric potential distribution as predicted by the 3D FEM model, and resulting droplet (\SI{90}{\nano\meter} radius, \SI{380}{\elementarycharge} charge) flight path for a \SI{0.7}{\nano\meter} branch offset. (c) Lateral (x) and vertical (z) landing position of the droplet derived experimentally from the attachment point of the secondary branch, as measured in SE micrographs, and comparison to predictions of the FEM model for different $r \cdot q$ products. Each experimental data point is averaged from 18 datasets obtained from 2 independent experiments (9 datasets each, see Figure~S2) and the error bars denote the resulting standard deviation. (d) Coefficient of determination (r\textsuperscript{2}) that reflects the fitting quality of the FEM model to the experimental data for different droplet radii and number of charges. The results highlight how the $r \cdot q$ product governs the droplet trajectory.}
	\label{fig:Figure_3}
\end{figure}

The autofocusing effect—i.e., the attraction of charged droplets to readily printed structures—represents both an opportunity and a challenge for the EHDRP process. 
On the one hand, it improves reliability in the fabrication of out-of-plane wire-like structures, such as pillars, by attracting charged droplets to the apex; it attenuates any imprecision in the positioning of the printing nozzle.
On the other hand, the resulting droplet deviation from a straight path complicates the fabrication of complex geometries, for which the use of non-trivial printing paths needs to be developed.
Here, we highlight this phenomenon by studying the droplet deflection in response to the presence of a \SI{2}{\micro\meter}-tall pillar as a function of the offset, i.e., the lateral nozzle displacement between the first pillar and the secondary branch.
The flight path—whose endpoint is revealed by the branch attachment point in the secondary electron (SE) micrographs—appears strongly modified by the presence of the first pillar (Figure~\ref{fig:Figure_3}(a)). This effect is amplified the smaller the offset between pillars.
For offsets of \SI{0.7}{\micro\meter} or smaller, the deflection is large enough that the droplet lands on the side of the first pillar. 
Considering the pillar width (\SI{190}{\nano\meter}), this observation implies a sideways deflection up to \SI{0.6}{\micro\meter} from the vertical path.

The map of the electric potential predicted by the FEM model shows how the small radius of curvature associated with the body and the tip of the pillar locally increases the magnitude of the electric field and bends the field lines towards it (Figure~\ref{fig:Figure_3}(b)), thereby leading to the substantial droplet deflection.
The FEM model can replicate the deflection behavior closely for all offsets investigated (Figure~\ref{fig:Figure_3}(c)), and the best fit is obtained for an $r \cdot q$ product of $\approx\SI{34.2}{\micro\meter\elementarycharge}$ (\SI{34(4)}{\micro\meter\elementarycharge} accounting for the error estimate in the experimental data).
In line with the computational results for droplet deflection by thin film electrodes, the $r \cdot q$ relationship holds here (Figure~\ref{fig:Figure_3}(d)) despite the fundamentally different configurations of droplet deflection caused by a protruding structure on the substrate.

\subsection{Relevance of the r$\cdot$q product to the flight dynamics of EHD droplets}

Predictions of the two aforementioned FEM models—electrostatic deflection by a thin-film electrode and autofocusing effect—reveal how the droplet flight path is closely bound to the product of its radius and charge: $r$ and $q$.
In both these configurations, droplets with identical $r \cdot q$ (or, in terms of mass, $m^{1/3} \cdot q$) follow similar trajectories (Figure~S3).
Quantitatively, the deviation from this relationship is small enough that, for the electrostatic deflection of droplets characterized by $r \cdot q = \SI{15.0}{\micro \meter \elementarycharge}$, increasing the radius by a factor 4 (from \SI{45}{} to \SI{180}{\nano \meter}, i.e., $64\times$ larger mass) leads to less than \SI{20}{\nano \meter} deviation in the landing position  (Figure~\ref{fig:Figure_2}(f)).
In comparison, a much smaller increase in the droplet size from \SI{45}{} to \SI{57}{\nano\meter}, while keeping the number of charges constant, already leads to a larger shift of \SI{30}{\nano \meter}.

Additionally, the trajectory deflection is inversely proportional to the $r \cdot q$ product.
A heavy or strongly charged droplet follows a straighter path (i.e., less impacted by the bending of the electric field lines) than a light or weakly charged droplet.
This result follows intuition with respect to the radius—a heavier droplet has more momentum and thus deflects less—but is more nuanced with respect to the charge.
While the lateral acceleration during the flight scales with stronger charge (Figure~S3(c)), a stronger charge also leads to an increase in the vertical acceleration (Figure~S3(b)) and, ultimately, less time allowed for lateral deflection.
Thus for a more strongly charged droplet, the reduced flight time allows for less lateral deflection despite the increase in lateral electrostatic force.

This result is particularly interesting given that the deflection of charges in response to an electric or magnetic field typically follows the mass-to-charge $m/q$ ratio and not $m^{1/3} \cdot q$.
In mass spectrometry measurements, $m/q$ is the ratio discriminating between species and is independent of the type of mass selection (sector field, time-of-flight, quadrupole filter).
Here, the observed difference in behavior arises from the aerodynamic drag force acting on the droplet in the model, which is negligible in mass spectrometry (see comparison in the absence of drag force, Figure~S4).
Consequently, droplet trajectories are sensitive to the dynamic viscosity of the surrounding medium which is determined by factors including gas composition, temperature, pressure. 
However, the $r \cdot q$ relationship holds independent of these conditions. 
As the magnitude of the drag force tends towards zero (Figure~S4(b))—for instance by reducing the gas pressure—all flight trajectories converge regardless of the $r \cdot q$ product. 
Our findings are taken and discussed in the context of the the experimental conditions reported: argon gas at atmospheric pressure and room temperature.

To support and understand the model's predictions with regards to the $r \cdot q$ relationship, a simple analytical model can be derived with the following assumptions:
\begin{enumerate}
    \item The space between the nozzle and the substrate is divided between two zones (acceleration, deflection) at a distance $h$ from the substrate.
    \item In the first zone, the droplet accelerates vertically due to the electric field, while the drag is neglected. The total voltage drop in this zone is denoted $\Delta V$.
    \item In the second zone, the droplet decelerates due to the drag force and is deflected laterally by a constant electric field $E_{x}$, while the vertical electric field is neglected.
    \item In the second zone, the lateral velocity is small compared to the vertical one (Figure~S3(b–c)), such that the lateral component of the drag force is neglected.
    \item The drag follows Stoke's law, $F=6 \pi \mu r v$, with $\mu$ as the dynamic viscosity, $r$ as the radius of the droplet, and $v$ as the droplet speed.
\end{enumerate}
With these assumption, the resulting deflection is calculated as $\Delta x = A \cdot r  \cdot q \cdot \ln^2(1-B/\sqrt{r \cdot q})$, with $A = \frac{1}{54 \pi}\frac{E_{x} \rho}{\mu^2}$, $B = 3\sqrt{\frac{3\pi}{2}}\frac{\mu h}{\sqrt{\rho \Delta V}}$, and $\rho$ as the solvent density.
In this relationship, $r$ and $q$ are only found in the form of their product $r \cdot q$.
Therefore, in accordance with the predictions of the FEM model, all droplets with identical $r \cdot q$ are equally impacted by the lateral electric field.
Moreover, the comparison between the experimental results and the predictions of the models (both numerical and analytical) only allows for the determination of its $r \cdot q$ product, not the individual values of $r$ or $q$.
In that regard, the predictions of the analytical and FEM models for the droplet deflection differ appreciably for the investigated range of size and charge.
For instance, the analytical model is not valid for $r \cdot q>B^2$ ($\approx\SI{80}{\micro \meter \elementarycharge}$, for larger $r \cdot q$, the argument of the logarithm is negative) which is not negligibly far from the value predicted by FEM.
Therefore, while the analytical model supports our phenomenological understanding and validates the FEM model, we will rely on numerical methods for a quantitative assessment of the droplet deflection.

The $r \cdot q$ products obtained from the two above-mentioned experimental approaches do not perfectly correspond: \SI{34(4)}{\micro \meter \elementarycharge} from the autofocusing experiment vs. \SI{7}{}–\SI{30}{\micro \meter \elementarycharge} from the experiment with a deflection electrode.
The origin of the discrepancy may arise from neglecting the role of solvent evaporation.
In experiments, the droplet mass decreases in flight due to evaporation and changes the $r \cdot q$ product.
Thus for longer flight times, the $r \cdot q$ product is on average smaller.
Since the nozzle distance was larger for the experiments with the deflection electrode (\SI{9}{} vs. \SI{7.5}{\micro \meter}), it may explain the smaller $r \cdot q$ derived from this approach.
Another source of error could be the absence of a dielectrophoretic force in the model.
Since this force is proportional to the gradient of the squared magnitude of the electrical field—which is different for the two configurations—it would alter the droplet path differently.
While the significance of the dielectrophoretic force in the droplet dynamics is still unclear, it is most likely negligible, as discussed in Section~S2 (Supporting Information).

The value derived from the autofocusing configuration appears more reliable for the two following reasons.
First, the estimated error in the SEM measurement is significantly smaller for this experiment (see error bars, Figure~\ref{fig:Figure_3}(c)).
Second, the FEM model of this configuration appears more sensitive to the $r \cdot q$ product as small changes in $r \cdot q$ lead to substantial variations in the prediction of the landing position.
The latter is particularly true around the transition between where the droplet lands on the pillar vs. the substrate (see Figure~\ref{fig:Figure_3}(c), branch offset between \SI{0.6}{} and \SI{1}{\micro \meter}).
Thus in the following analysis, we will assume an $r \cdot q$ value of \SI{34(4)}{\micro \meter \elementarycharge}.

One can further narrow down this estimation of $r \cdot q$ by considering the Rayleigh limit\cite{Rayleigh1882}, $n e = 8 \pi \sqrt{\epsilon_{0} \gamma r^3}$, which sets the maximal number ($n$) of elementary charges ($e$) a droplet of radius ($r$) and surface tension ($\gamma$) can sustain without Coulomb explosion.
Taking \SI{34}{\micro \meter \elementarycharge} for $r \cdot q$ and \SI{28.66}{\milli \newton \per \meter} for $\gamma$\cite{Haynes2016}, the Rayleigh limit is reached for a \SI{45}{\nano \meter} radius, which corresponds to a charge of \SI{755}{\elementarycharge}.
Therefore, the droplet characteristics have to satisfy both, $r \cdot q = \SI{34}{\micro\meter\elementarycharge}$ and $r>\SI{45}{\nano\meter}$.
For the EHD process, it is generally assumed that the droplet size is on the order of the feature size, which, for a pillar of \SI{190}{\nano\meter}, would imply a radius of $\approx\SI{95}{\nano\meter}$ and thus a charge of $\approx\SI{358}{\elementarycharge}$.

Although the exact droplet size and charge remain uncertain, the results of the previous sections showed that the $r \cdot q$ product is sufficient to predict the flight behavior.

\subsection{Printing paths compensating for the autofocusing effect}

\begin{figure}[htbp]
    \centering
    \includegraphics[width=17.78cm]{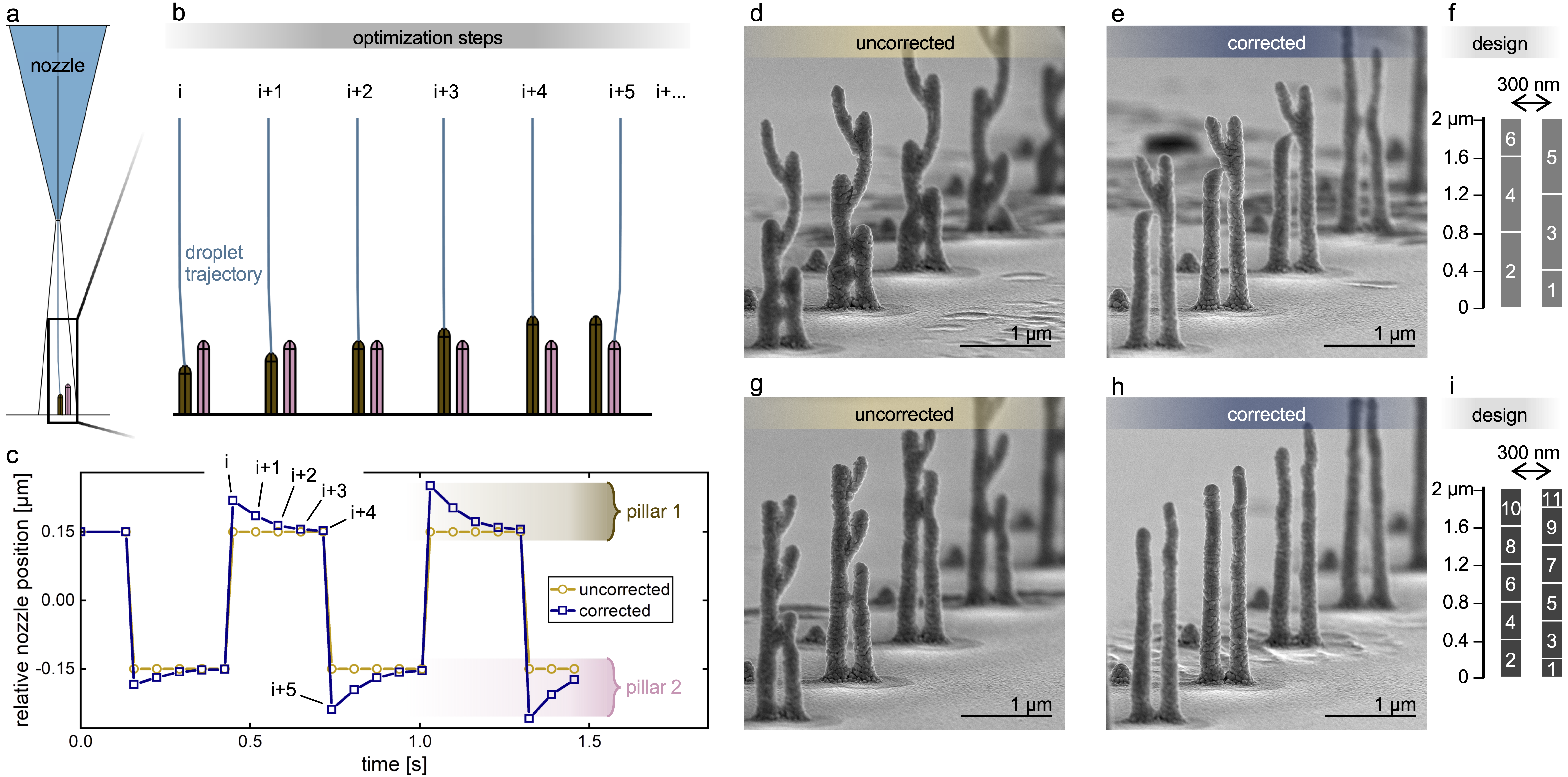}
	\caption{\textbf{Printing path optimization for compensation of the autofocusing effect.} (a) Overview of the geometrical configuration of the FEM model. (b) Flight path (blue) after optimization of the nozzle positioning for a selection of growth steps. All trajectory endpoints are correctly situated at the apex of the growing pillar. (c) Nozzle positions with and without correction as a function of printing time for \SI{800}{\nano\meter}-long pillar segments, considering a \SI{3}{\micro\meter\per\second} vertical growth rate. SE micrographs displaying the experimental outcomes for (d–f) \SI{800}{\nano\meter} and (g–i) \SI{400}{\nano\meter} segment length. The corrected paths enable the deposition of nanopillars with a high degree of geometrical conformity (e,h), while appreciable divergence is observed in the absence of correction (d,g). Imaging tilt: 80°.}
	\label{fig:Figure_4}
\end{figure}

Using the $r \cdot q$ calculated from the autofocusing experiments, we transform the FEM model into an optimization tool in which the lateral nozzle position is iteratively adapted to reduce the absolute distance between the calculated and the target landing position. 
This enables a corrected nozzle path that compensates for the autofocusing effect.
Note that we choose the $r \cdot q$ derived from the autofocusing experiments, \SI{34}{\micro\meter\elementarycharge}, firstly for the reasons discussed in the previous paragraph, and secondly since the model should correct for this exact phenomenon.
As a proof of principle, we derive the optimal printing path for a simple geometry which, despite its apparent ease, represents a challenge in the absence of print corrections: two closely spaced nanopillars.
These are designed to be \SI{2}{\micro\meter} in height and spaced by \SI{300}{\nano\meter}, resulting in a gap of half the pillar width (assuming \SI{200}{\nano\meter}-thick cylinders).
As visible in the left-most micrograph of Figure~\ref{fig:Figure_3}(a), this geometry represents a challenge that cannot be achieved by printing the first pillar and afterward the second—without correction, the second pillar grows from the tip of the first.
One simplistic approach that decreases the autofocusing effect is layer-by-layer printing. 
Thus, to provide a fair comparison between this uncorrected printing strategy and the FEM-derived path corrections, the two pillars are printed in a layer-by-layer fashion, with each pillar segment being either \SI{400}{} or \SI{800}{\nano\meter} in height.

The nozzle positions optimized by the FEM model and corresponding droplet flight path are shown for a subset of growth steps in Figure~\ref{fig:Figure_4}(a–c).
The correction is stronger—and therefore the more crucial—the taller the non-growing pillar is compared to the growing one.
The magnitude of the correction also increases with the overall height of the two nanopillars (Figure~\ref{fig:Figure_4}(c)), which stresses the importance of the correction for larger and more complex structures.
The results also highlight the limitation of the layer-by-layer approach on reducing the autofocusing effect.
Even in the extreme case of an infinitely thin layers, i.e., the two pillars have the same height during the whole growth process, a significant correction still applies: \SI{14}{\nano\meter} for step \textit{i+2} in Figure~\ref{fig:Figure_4}(b–c) and  even \SI{24}{\nano\meter} when the 2 pillars reach \SI{2}{\micro\meter} in height (last growth step).
This correction seems small but without it, the error in the pillar apex position would increase for each new layer, and the final print outcome would not appear correct.
This illustrates the limited potential for a layer-by-layer printing scheme to decreasing the autofocusing effect and emphasizes the necessity of FEM-calculated printing path corrections.

The experimental results further corroborate the benefit of this approach.
Pillars fabricated using a corrected path (Figure~\ref{fig:Figure_4}(e,h)) are considerably closer to the design geometry than the uncorrected ones (d,g), for both \SI{800}{\nano\meter} segments steps (f) and \SI{400}{\nano\meter} segments steps (i) (see also similar results from an independent experiment in Figure~S5).
The small error visible at the top of the corrected pillars with large segments steps (e) may have two possible origins.
First, the pillars shown are slightly higher than designed: \SI{2.3}{\micro\meter} instead of \SI{2}{\micro\meter} (growth rate was $\approx\SI{3.45}{\micro\meter\per\second}$ instead of the expected \SI{3}{\micro\meter\per\second}).
Consequently, the antenna effect was stronger while printing than in the simulation, and may explain the increased deflection observed.
Second, the top of the pillar is thinner than the bottom—and thus grew faster (assuming a constant volumetric deposition rate)—resulting in an increased droplet attraction.
In any case, the defect remains minor in comparison to the these of the uncorrected geometry.
Thus, the results demonstrate the potential of FEM to bridge the gap between basic geometries achieved by trial and error—the current state of the art in EHD printing—and the fabrication of arbitrarily complex 3D shapes.

\subsection{Towards arbitrary geometries}

\begin{figure}[htbp]
    \centering
    \includegraphics[width=17.78cm]{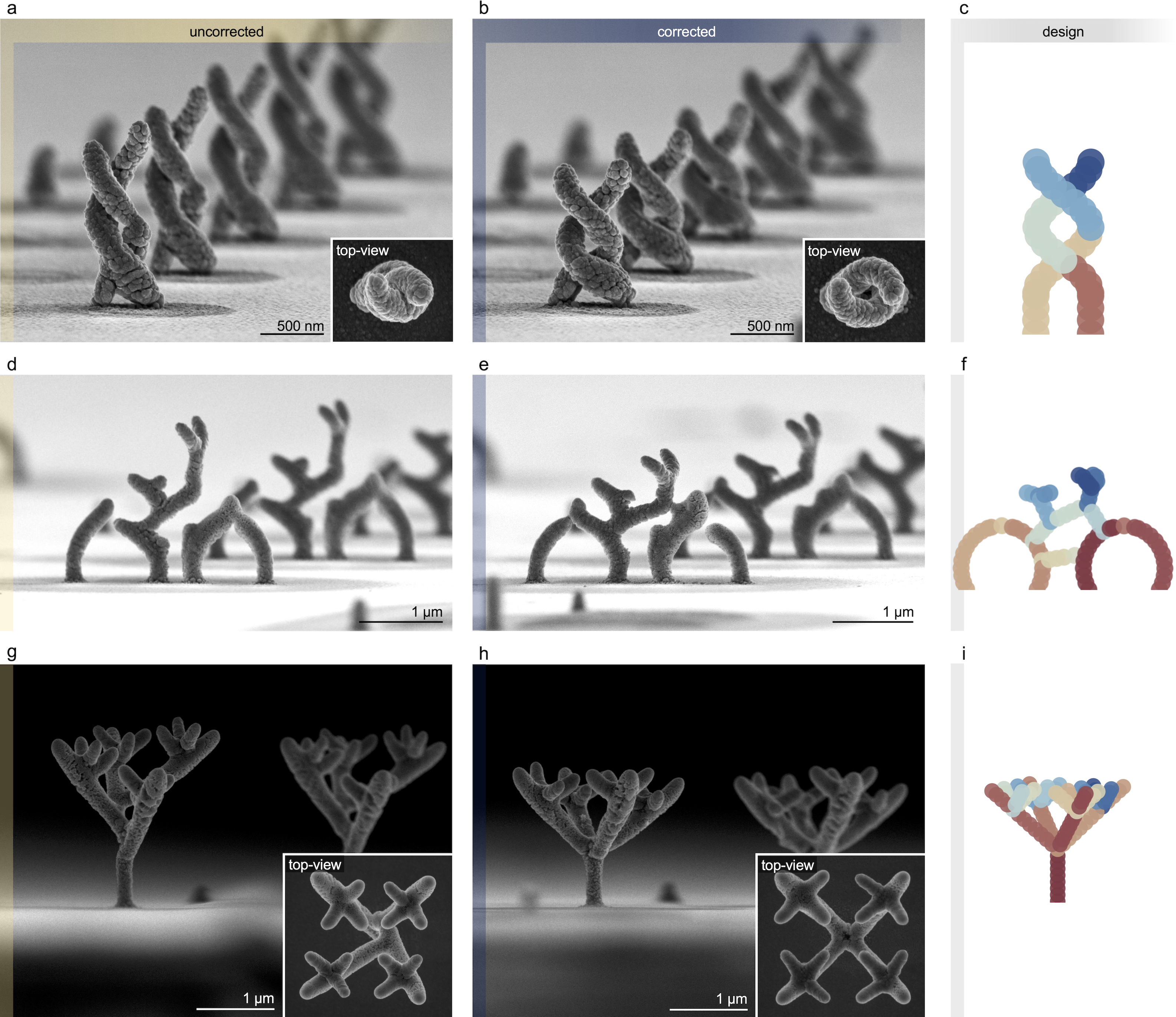}
	\caption{\textbf{Unlocking the fabrication of arbitrary shapes.} Nanoscale 3D geometries deposited by EHDRP (a,d,g) without and (b,e,h) with autofocusing-corrected printing paths. (c,f,i) The 3D models—(c) double helix, (f) bicycle, (i) tree-like network—are discretized in spherical building blocks. From red to blue, the spheres' color denotes the order in which each segment was printed. Imaging tilt: (a–b) 85°, (d–e) 86.5°, (g–h) 89.5°.}
	\label{fig:Figure_5}
\end{figure}

To generalize this approach to the fabrication of arbitrary shapes, the 3D models are discretized into printing steps, each corresponding to an added spherical voxel. 
For each printing step, i.e., building block, the FEM model constructs the geometry as the union of the previous spheres and then optimizes the nozzle position accordingly.
The experimental outcomes for a double helix, a bicycle, and a tree-like network with and without corrected printing paths are visible in Figure~\ref{fig:Figure_5}.
In absence of correction, the topology of the fabricated structures is not guaranteed: the two strands of the double-helix (a) collapse on each other, and the four main branches of the nano-trees (g) do not emanate from a unique point.
Conversely, the corrected nano-objects display an excellent geometrical conformity with the model and good reproducibility.
Thus, the use of autofocusing-corrected printing paths unlocks the fabrication of complex 3D shapes by EHD printing and opens new prospects for this additive manufacturing technique in various field of nanotechnologies.

These results illustrate both the success of model-based corrections and some of the fabrication challenges linked to EHD 3D printing.
First, the voxels sitting on the substrate require additional residence time for an equivalent volume of deposited material as compared to subsequent layers. 
This leads to shorter pedestals for the spiral's strands (Figure~\ref{fig:Figure_5}(a–c)) and a larger cropping of the lower parts of the bicycle wheels (Figure~\ref{fig:Figure_5}(d–f)).
This effect is likely related to a spreading of the copper nuclei on the substrate for the first landing droplets, and may be corrected for by investigating the delayed time for the pillar nucleation.
Second, the bicycle frame illustrates the difficulty to fabricate overhanging structures with beam angle approaching 0° from the horizontal. 
In this example, the top tube at 25° above horizontal (between saddle and handlebars) was fabricated successfully, while the 9° down tube (which has been simplified here into a tube connecting the wheels) was not.
Although 0° beams have been reported \cite{Reiser2019}, their fabrication requires a precise match between the growth rate and the nozzle movement, and thus any imprecision in the assumed growth rate can be detrimental.
Third, bridges—such as the bicycle wheels—can be printed successfully but shows limited reproducibility.
This topology is particularly challenging because the two sides of the wheel need to have the exact same height, otherwise the droplets ejected for the final link are attracted by the higher side and a gap remains.

\section{Conclusion}

We present two experimental approaches to study the deflection of charged droplets involved in the EHDRP 3D printing process.
First, a voltage signal applied to external deflection electrodes directly controls the trajectory of EHD droplets and allows for their precise positioning in the fabrication of 2.5D structures.
Second, alteration to the droplet's landing position due to a nearby nanopillar is observed to quantify the autofocusing effect directly and highlights the related challenges in achieving high geometrical complexity by EHD.
The reported FEM simulations agree with both experimental approaches and unveil that the product of droplet size and charge predominantly governs its flight dynamics.
In the current study, this quantity ($r \cdot q$) lies around \SI{34}{\micro\meter\elementarycharge} and likely corresponds to a droplet radius of $\approx\SI{95}{\nano\meter}$ and a charge of $\approx \SI{358}{\elementarycharge}$.
Finally, we show how the printing paths can be optimized by an FEM model to compensate for the electrostatic deflection of the charged droplets.
This strategy unlocks the deposition of complex 3D nano-objects that are not achievable otherwise, and unleashes the full potential of EHD 3D printing in the field of microfabrication.

\section{Methods}

\subsection{Printing procedure and setup}

The printing nozzles were prepared with a pipette puller (P-2000, Sutter Instrument) from  quartz capillaries (QF100-70-15, Sutter Instrument).
The pulling parameters were: Cycle 1: Heat 810, Filament 5, Delay 128, Pull 50, Velocity 30; Cycle 2: Heat 700, Filament 4, Delay 130, Pull 75, Velocity 50, leading to aperture diameters of 150–\SI{220}{\nano\meter}.
The nanopipettes were filled with acetonitrile (Optima, Fisher Chemical) using a glass syringe (Gastight \#1010, Hamilton) equipped with a \SI{20}{\nano\meter} syringe-filter (Anotop 10, GE Whatman).
Cu wires used as sacrificial anodes (\SI{99.9999}{\percent}, \SI{0.25}{\milli\meter}-thick, Alfa Aesar) were etched in \SI{65}{\percent} concentrated nitric acid (Sigma Aldrich) for \SI{10}{\second}, rinsed in water (Optima LC/MS grade, Fisher Chemical), stored in ethanol (EMSURE, absolute for analysis, Supelco, Merck) and finally inserted into the nanopipettes.
The printing process was conducted in an inert Ar atmosphere with $<$100~ppm \ce{O2}, as monitored by a Module ISM-3 oxygen sensor (PBI Dansensor).
Substrates used for printing were \ce{Si} wafers ((100), SiMat, with \SI{50}{\nano\meter} diffusion barriers of \ce{SiO2} and \ce{Si3N4}) coated by DC magnetron sputtering (PVD Products Inc.) with a \SI{3}{\nano\meter} \ce{Ti} and a \SI{80}{\nano\meter} Au film.
The thin-film electrodes were made by standard lithography process, with a patterning step performed by laser writing (DWL66+, Heidelberg).
The steering setup with electrical wiring is visible in Figure~S6.
The nanopositioning stage (QNP piezo stage with Ensemble\textsuperscript{\textregistered} controller, Aerotech) and the printing voltage provided by a power source (B2962, Keysight) were controller with a custom Matlab script.
The printing voltage was set to \SI{110}{\volt} in all experiments.
The ramping profiles of the piezo stage between the autofocusing-corrected waypoints were performed by a script written in the AeroBasic\textsuperscript{\texttrademark} programming language and running on the stage controller.

\subsection{Microscopy}

The samples were imaged with a FEG-SEM (Magellan 400, FEI, USA) and with a FIB-SEM station (NVision40, Zeiss, Germany), both operating with in-lens mode with an acceleration voltage of \SI{5}{\kilo\volt} for the electron source.

\subsection{FEM model}

All 3D numerical models were built in COMSOL Multiphysics\textsuperscript{\textregistered} software.
The Electrostatics module computed the electric potential distribution for a printing potential applied at the capillary surface and a grounded substrate.
The droplet trajectory was calculated thereafter using the Charged Particle Tracing module, comprising an electrostatic force and a drag force as defined by Feng \textit{et al.} for the motion of a viscous sphere subjected to interfacial slip\cite{Feng2012}.
The correction for a viscous sphere—accounting for the existence of flows within the droplet—is preferable when the particle viscosity is not substantially larger than the surrounding fluid (here, the viscosity ratio is only $\approx15$), while the interfacial slip correction is justified by the large Knudsen number in the present study ($\approx1$, defined as the mean free path of gas molecules over particle radius).
The nozzle positions that best compensate for the autofocusing effect were determined using the derivative-free BOBYQA (Bound Optimization BY Quadratic Approximation) method, that iteratively adapts the model geometry to reduce the deviation between the computed and the desired landing position.
The double pillars geometry in Figure~\ref{fig:Figure_4} was made of two cylinders with ellipsoid ends and having variable total heights.
The arbitrary shapes in Figure~\ref{fig:Figure_5} are wire-like structures modeled and discretized in Blender software.
The file conversion from Blender to COMSOL Multiphysics\textsuperscript{\textregistered} and ultimately to the 3D-printer software was managed by a custom Wolfram Mathematica script.

\begin{acknowledgement}

Electron microscopy analysis was performed at ScopeM, the microscopy platform of ETH Z\"urich.
Sputter deposition of thin films were performed at FIRST, the clean room facilities of ETH Z\"urich.
The authors thank Carmen Lauener for the fabrication of the substrates with thin-film electrodes.
The authors are grateful to Jelena Wohlwend and Rebecca Gallivan for their excellent and constructive feedback on the manuscript.
The work was funded by the SNF Grant 200021\textunderscore188491.

\end{acknowledgement}

\bibliography{manuscript}

\end{document}